\documentclass[12pt]{iopart}

\usepackage{graphicx}
\usepackage{iopams}
\newcommand{\Cee}{\ensuremath{\mathcal{C}}}

\begin{document}

%%%%%%%%%%%%%%% TITLE AREA %%%%%%%%%%%%%%% 
\title{The Grand-Canonical Asymmetric Exclusion Process and the One-Transit Walk}
\date{February 2004}

\author{R.\ A.\ Blythe}
\address{Department of Physics and Astronomy, University of Manchester,
Manchester, M13 9PL, England}

\author{W.\ Janke}
\address{Institut f\"ur Theoretische Physik, Universit\"at Leipzig,
Augustusplatz 10/11, 04109 Leipzig, Germany}

\author{D.\ A.\ Johnston}
\address{School of Mathematical and Computer Sciences,
Heriot-Watt University, Riccarton, Edinburgh EH14 4AS, Scotland}

\author{R.\ Kenna}
\address{School of Mathematical and Information Sciences,
Coventry University, Coventry, CV1 5FB, England}

\begin{abstract}
The one-dimensional Asymmetric Exclusion Process (ASEP) is a paradigm
for nonequilibrium dynamics, in particular driven diffusive processes.
It is usually considered in a canonical ensemble in which the number of sites 
is fixed.  We observe that the {\it grand}-canonical partition function for the 
ASEP is remarkably simple.  It allows a simple direct derivation of the
asymptotics of the canonical normalization in various phases
and of the correspondence with One-Transit Walks recently observed
by Brak~{\em et~al.}
\end{abstract}


\maketitle

%%%%%%%%%%%%%%% MAIN TEXT %%%%%%%%%%%%%%% 

\section{Introduction}
Driven physical systems that are out of equilibrium are characterized by steady states 
in which currents may flow.  Examples of such systems are provided by the kinetics of 
biopolymerization \cite{MGP}, transport across membranes \cite{CLprl} and traffic flow 
\cite{CSS}.  The existence of a macroscopic steady-state current has an important 
consequence for the microscopic modeling of such systems.  Dynamics which are microscopically 
reversible (i.e.\ related to an energy function through the detailed balance condition) do not, 
by definition, admit macroscopic currents.  Hence to understand nonequilibrium---and, in 
particular, driven---systems, it is useful to study closely microscopically 
\textit{irreversible\/} processes.

There is, at present, no unified statistical mechanical theory underlying 
nonequilibrium systems, and this presents a major barrier against progress 
towards a general, deep understanding thereof.  Nevertheless, there is a 
class of one-dimensional driven-diffusive systems---typified by the 
Asymmetric Exclusion Process (ASEP) defined below---that have been 
exactly solved and shown to exhibit nontrivial steady-state phenomena 
such as phase transitions \cite{DEHP}, spontaneous symmetry breaking 
\cite{EFGMprl} and jamming \cite{OEC}.  More recently, there have been 
some advances in extending notions from equilibrium statistical mechanics 
to nonequilibrium steady states.  These include the Lee-Yang theory of 
partition-function zeros and phase transitions \cite{Arndt,BEprl,BEbjp} and, 
in a separate development, the derivation of nonequilibrium free energy 
functionals \cite{DLSprl}.

It might be that the success of these approaches is due to the ASEP's steady 
state being related to certain types of random walks.  This is because the 
latter are objects that are meaningful in the framework of equilibrium 
statistical physics---see, e.g., \cite{JvR} for a comprehensive overview.  
The connection between the steady state of the ASEP and \textit{One-Transit Walks} was investigated in \cite{Brak1,Brak2}.  In particular, it was shown that 
the transfer-matrix representation of the  partition function can be interpreted 
in the framework of a similar formalism which proved instrumental for the exact 
solution of the ASEP \cite{DEHP}.  In the present work, we show that the connection to random walks is transparent in an ensemble in which the system size of the ASEP fluctuates.  In the equilibrium picture, it is the lengths of the random walks that fluctuate, and so we deem the ensemble \textit{grand canonical}.  In the course of this work we will show that the grand-canonical partition function has a surprisingly compact form, but one that nevertheless elegantly encodes the phase behavior of the ASEP in the canonical (fixed system size) ensemble.  We conclude our work with some speculation as to the existence of more elementary solutions of the ASEP and related models based on the compactness of this formula.

We begin by recalling the deceptively simple dynamics of the ASEP with open 
boundaries, indicated in Fig.~1.  Particles are introduced at a rate $\alpha$ 
per unit time at the start of an $N$-site chain and leave at a rate $\beta$ 
at the other end. They can hop with probability one per unit time to the 
right if the space is empty, otherwise they remain stationary.
\begin{figure}
\begin{center}
\includegraphics[scale=0.6]{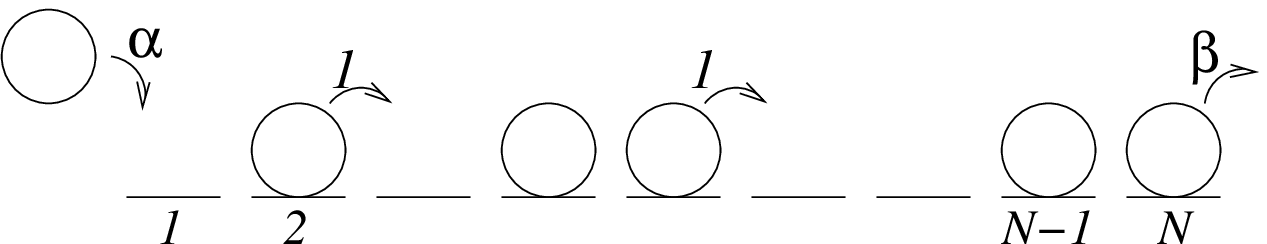}
\caption{\label{asep-fig}Dynamics of the ASEP.  The arrow
labels indicate the rates at which the corresponding transitions occur;
site labels are also indicated.}
\end{center}
\end{figure}
Although the model is a nonequilibrium construct, defined purely in terms 
of microscopically irreversible transitions, it still displays 
steady-state solutions of various sorts.  A key quantity in describing 
the behavior of the model is the current of particles, which acts as an 
order parameter in the different phases.

The model displays a phase diagram that possesses both first- and 
second-order phase transitions.
\begin{figure}
\begin{center}
\includegraphics[scale=0.60]{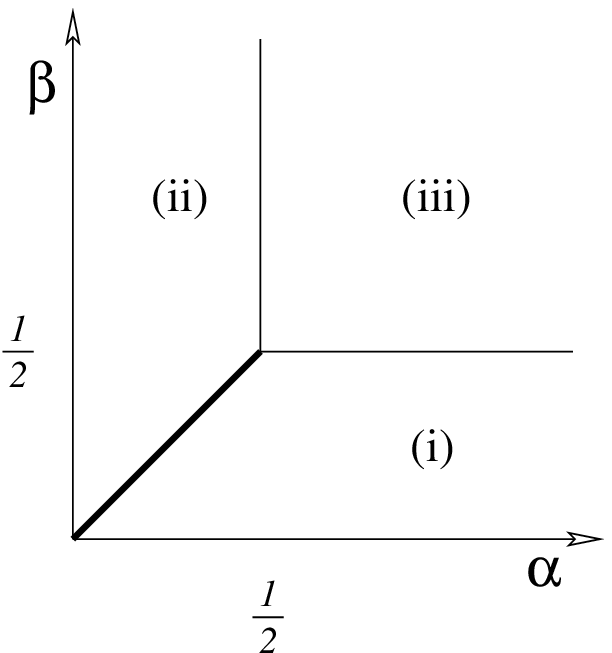}
\caption{\label{Jphase-fig}Phase diagram of the ASEP.}
\end{center}
\end{figure}
In Fig.~2, phase (i) is a high-density phase with a density profile that 
decays
exponentially towards the right boundary and  has a current $J=\beta(1-\beta)$. 
Phase (ii) on the other hand is a low-density phase that has an exponential 
decay in the density profile from the {\it left} boundary and a current
$J=\alpha(1-\alpha)$.  The line $\alpha = \beta < \frac{1}{2}$ is a first-order 
transition line at which the current exhibits a discontinuity in its first 
derivative.  Finally in phase (iii), the current assumes a maximal constant 
value of $J = \frac{1}{4}$.  Since $J$ has a discontinuity in its second
derivative passing from phase (iii) to either (i) or (ii) the horizontal and 
vertical boundaries are second-order transition lines.

It is at present unknown whether one can define for general nonequilibrium 
steady states a partition function from which all thermodynamic information 
can be obtained.  Nevertheless it is known that a quantity constructed by 
summing the statistical weights (unnormalized probabilities) $f(\Cee)$ 
over all configurations $\Cee$ in the steady state of the ASEP does at 
least capture the model's phase behavior (by applying, e.g., a Lee-Yang 
analysis \cite{BEprl}).  
Denoting this analog of the partition function by $Z$, we have
\begin{equation}
\label{Zgen}
Z = \sum_{\Cee} f(\Cee)\;,
\end{equation}
where the weights $f(\Cee)$ are defined by the microscopic transition 
rates $W (\Cee \to \Cee^\prime)$ in the model and the requirement that 
probability fluxes in the steady state balance:
\begin{equation}
\label{steadystate}
\sum_{\Cee^\prime \ne \Cee} \left[ f(\Cee^\prime) W(\Cee^\prime \to
\Cee) - f(\Cee) W(\Cee \to \Cee^\prime) \right] = 0 \;.
\end{equation}
The probability of a configuration $\Cee$ is then given by $P(\Cee) = f(\Cee)/Z$.  
Since  the weights, $f(\Cee)$, are fixed only up to 
an overall scale by (\ref{steadystate}), 
one can ensure that $Z$ is uniquely defined by, for example, 
insisting that it is polynomial in the transition rates $W(\Cee \to \Cee^\prime)$ 
with any factors common to all weights $f(\Cee)$ removed \cite{RABthesis,Brak2}.

The tour de force in \cite{DEHP} gives exact formul\ae\ for these weights for an 
ASEP with $N$ sites.  The resulting expression for the canonical partition 
function, $Z_N$, reads
\begin{equation}
\label{asepZ}
Z_N = \sum_{p=1}^{N} \frac{p(2N-1-p)!}{N!(N-p)!}  \frac{
(1/\beta)^{p+1} - (1/\alpha)^{p+1} }{ (1/\beta)-(1/\alpha) } \;.
\end{equation}
The current is given by $J = Z_{N-1}/Z_N$ and various correlation functions
can also be obtained \cite{DEHP}.  For later convenience we note that 
combinatorial factor 
appearing in (\ref{asepZ}) is given by the Ballot numbers $B_{N,p}$
\begin{equation}
\label{ballot}
B_{N,p} = \frac{p(2N-1-p)!}{N!(N-p)!} \;,
\end{equation}
so that
\begin{eqnarray}
Z_N &=&  \sum_{p=1}^{N} B_{N,p}   
\frac{ (1/\beta)^{p+1} - (1/\alpha)^{p+1} }{ (1/\beta)-(1/\alpha) }  \nonumber \\ 
\label{Zcanonq} &=& \sum_{p=1}^{N} B_{N,p} \sum_{q=0}^p
\frac{1}{{\alpha}^q} \frac{1}{{\beta}^{p-q}} \;.
\end{eqnarray}

Ballot numbers arise in the enumeration of \textit{Dyck paths} which are 
walks on a diagonally-rotated square lattice with the following properties: 
(a) they are constrained to lie above or on the $x$-axis; (b) they can only 
move in the north-east or south-east directions; and (c) they start \textit{and} 
end on the $x$-axis.  The quantity $B_{N,p}$ is the number of Dyck paths with 
length $2N$ and $p$ returns to the $x$-axis, including the final return that 
is always present \cite{JvR}.

\begin{figure}
\begin{center}
\includegraphics[scale=0.35]{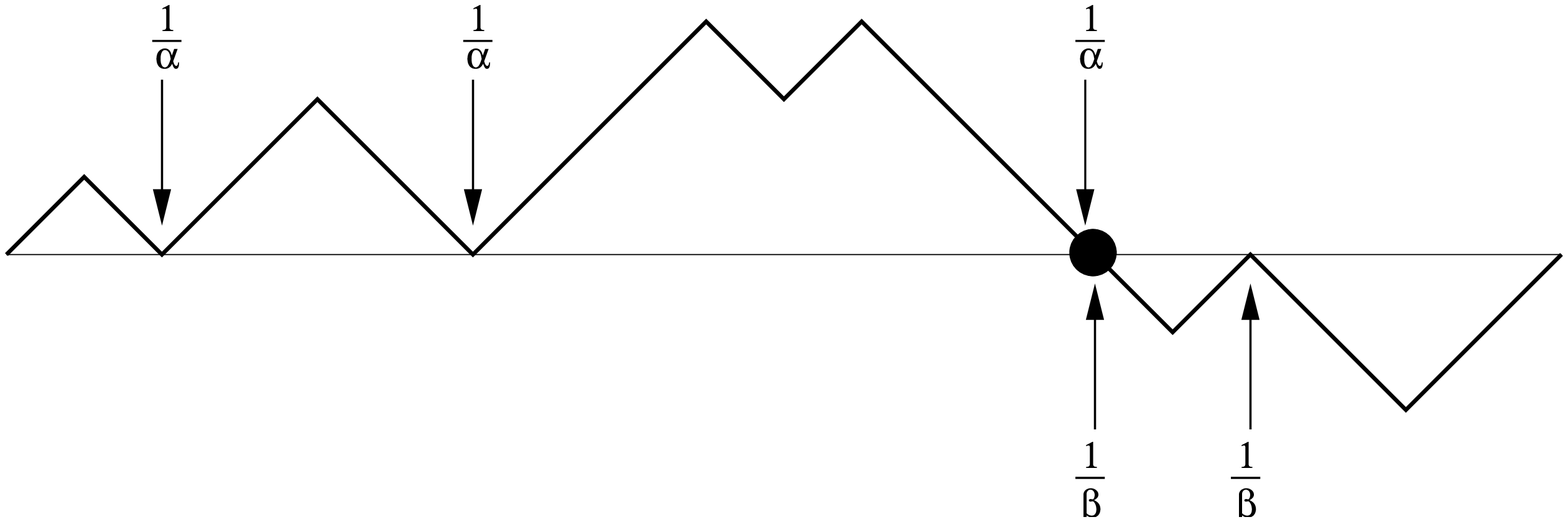}
\end{center}
\caption{\label{OTW-fig}
A One-Transit Walk obtained by composing two Dyck paths. Each contact with 
the $x$-axis from above,  apart from the start, gives a factor $1/\alpha$ and 
each contact from below, apart from the end, gives a factor of $1/\beta$.}
\end{figure}

With this in mind, (\ref{Zcanonq}) can be interpreted as a weighted sum 
over paths as follows.  Take a Dyck path from $(0,0)$ to $(2N,0)$ with $p$ 
returns, and from the $q^{\mathrm{th}}$ return onwards ($q=0,1,\ldots, p$) 
reflect the path about the $x$-axis.  The resulting construct is a 
\textit{One-Transit Walk}, an example of which is shown in 
Fig.~\ref{OTW-fig}.  The weight, $\alpha^{-q} \beta^{-(p-q)}$, associated 
with the walk can be gleaned by assigning a factor $1/ z_1 = 1/ \alpha$ 
to each contact with the $x$-axis from above, excluding an initial upward 
step, and similarly a factor $1/z_2 = 1/\beta$ to each contact from below, 
excluding a final upward step.  Note that initial or final downward steps 
\textit{do} contribute factors of $1/z_2$ and $1/z_1$ respectively since 
these correspond to the cases $q=0$ and $q=p$.

A slightly different scheme is that implemented in \cite{Brak1}.  There, 
a fugacity, $z_1$, is assigned to each down step and $z_2$ to each up step, 
apart from those ending on the $x$-axis. The partition function of this 
model is
\begin{equation}
\label{part-OTW}
\hat Z_N = (z_1 z_2)^N \sum_{p=0}^N B_{N,p} \sum_{q=0}^p z_1^{-q} z_2^{-p+q}
\;,
\end{equation}
which has the same form as $Z_N$ up to the analytic factor of $(z_1 z_2)^N$.  
This prefactor, which brings (\ref{asepZ}) into the standard form proposed 
in \cite{RABthesis,Brak2}, has no bearing on the critical behavior.

\section{A Grand Canonical ASEP}
We now switch to an alternative viewpoint of the ASEP which allows both 
the phase behavior and the connection to One-Transit Walks to be seen in 
a very straightforward way.  The key step is to relax the constraint of 
a fixed system size $N$.  To this end we introduce a fugacity, $z$, which 
is conjugate to $N$ and define a grand-canonical partition function, 
$\mathcal{Z}$, as
\begin{equation}
\label{GC}
{\cal  Z} (z) = \sum_{N=0}^\infty Z_N z^N \;.
\end{equation}
After substituting in the expression (\ref{asepZ}) for $Z_N$ and changing 
the order of summation, one can use \cite{Wilf}
\begin{equation}
\left( { 1 - \sqrt{1 - 4 z } \over 2 z } \right)^p =  \sum_{N=0}^\infty
{ p ( 2 N + p - 1 ) ! \over p! ( N + p )!} z^N \;,
\end{equation}
to give the following remarkably simple result for $\cal Z$:
\begin{equation}
\label{GCASEP}
{\cal  Z} (z) = { \alpha \beta \over ( x(z) - \alpha ) ( x(z) - \beta )} \;.
\end{equation}
In this expression $x(z)$ is the generating function for Catalan numbers,
\begin{equation}
\label{xz}
 x(z) = \frac{1}{2} \left( 1 - \sqrt{1 - 4 z} \right) \;,
\end{equation}
which crops up in numerous combinatorial contexts \cite{Comtet}, such as 
counting well-bracketed words or bifurcating trees. Writing ${\cal Z}(z)$
explicitly in terms of $z$, we have
\begin{equation}
\label{GCASEPz}
{\cal  Z} (z) = { 4 \alpha \beta \over    (  1 - 2 \alpha - \sqrt{ 1 - 4z }  )  
( 1 - 2 \beta -  \sqrt{ 1 - 4z }) }
\;,
\end{equation}
or, when $\alpha=\beta$,
\begin{equation}
\label{GCASEPz2}
{\cal  Z} (z) = \left[  {2 \alpha \over 1 - 2 \alpha - \sqrt{ 1 - 4z } }\right]^2 \;.
\end{equation}
One can alternatively use the Lagrange inversion formula with $x=z/(1-x)$ to show that expanding ${\cal Z} (x ( z ))$ also gives the desired result in (\ref{GC}).

We remark that the expression (\ref{GCASEPz}) has recently been derived in two distinct contexts.  First, it turns out to be instrumental in determining the joint distribution of density and current in the ASEP \cite{Dep2}.  To obtain the grand-canonical expression, a novel ``relaxed'' operator algebra method was invoked \cite{Dep1}.  Meanwhile, in a version of the ASEP in which the dynamics allow the system size to fluctuate (thereby giving physical meaning to the grand-canonical ensemble), an expression similar to, but more general than (\ref{GCASEPz}) arises by determining a quantity related to the resolvent of a matrix \cite{Heilmann}.

Standard results for the asymptotics of expansions of functions \cite{FS}
allow us to read off the known large $N$ behavior of $Z_N$ \cite{DEHP}
directly from (\ref{GCASEPz}) and (\ref{GCASEPz2}). 
The large $N$ behavior of $Z_N$ is dominated by whichever 
pole, $x \sim \alpha$ or $x \sim \beta$, is closest to the origin, or, if 
$\alpha, \beta >1/2$, by the square root singularity in $x(z)$ itself. 
The first two cases correspond to regions (i) and (ii) in the phase 
diagram (Fig.~2) and the latter to region (iii). In the latter case,
with $\alpha < \beta$ 
for example, the square root singularity in $\sqrt{1 - 4 z}$ in 
(\ref{GCASEPz}) gives rise to a $4^N / N^{3/2}$ asymptotic behavior and 
substituting in $z = 1/4$ picks up the coefficient
\begin{equation}
\label{asymp3}
Z_N \sim \frac{4^N}{ \pi^{1/2} N^{3/2} } 
{ \alpha \beta \over ( \beta - \alpha )}
\left[ { 1 \over (  2 \alpha - 1 )^2 } - { 1 \over (  2 \beta - 1)^2 }\right] \;,
\end{equation}
in perfect agreement with \cite{DEHP}.  The  $\alpha = \beta$ result in region 
(iii) follows in a similar fashion from (\ref{GCASEPz2}).

In regions (i) and (ii), on the other hand, it is the pole
terms which lead to the dominant asymptotic behavior. For (ii), 
the pole at $x = \beta, z= \beta ( 1 -\beta)$ is dominant since 
$\alpha > \beta$ and the square root singularity at $z=1/4$
has not yet been reached. 
In this case a $1/ ( \beta ( 1 - \beta ) )^{N}$ behavior results from
the pole term, giving
\begin{equation}
\label{asymp2}
Z_N \sim {\alpha ( 1 - 2 \beta )\over (\alpha - \beta) ( 1 - \beta) } { 1 \over ( \beta
( 1 - \beta ) )^{N}} \;.
\end{equation}
Finally, on the second-order boundary lines ($\alpha=1/2$, $\beta>1/2$) and 
($\beta=1/2$, $\alpha>1/2$) the singularity is of the form $1/\sqrt{1 - 4 z}$ 
and gives an asymptotic behavior of the form $4^N/N^{1/2}$. For ($\alpha=1/2$, $\beta>1/2$), 
for instance,
\begin{equation}
\label{asymp3a}
Z_N \sim \frac{4^N}{ \pi^{1/2} N^{1/2} } {2 \beta \over 2 \beta  -1 } \;.
\end{equation}

There is a close link between the grand-canonical partition function for the ASEP and the random walk picture, as we now show.  The generating function for Dyck walks with a fugacity $w$ for each contact with the $x$-axis is known to be given by \cite{JvR}
\begin{equation}
\label{Dyck}
G_D (w, z) = \frac{ w} { 1 - w ( 1 - \sqrt{1 - 4 z}) /2 }\;,
\end{equation}
in which $z^{1/2}$ is the fugacity for each step.  Comparison with (\ref{GCASEPz}) reveals that $\mathcal{Z}$ can be written as
\begin{equation}
\label{2Dycks}
{\cal  Z} (z) = \alpha \beta \; G_D ( 1/ \alpha, z ) \; G_D ( 1 /\beta , z) \;.
\end{equation}
The product form of this expression implies that the grand-canonical ensemble of 
paths comprises two weakly-interacting Dyck paths of fluctuating length, one with a contact fugacity $w=1/\alpha$ and the other with contact fugacity $w=1/\beta$. 
One way in which such paths can be realised is as One-Transit Walks as shown in 
Fig.~\ref{OTW-fig}.  Then, the $\alpha \beta$ prefactor simply takes account of 
the unweighted initial and final steps.  Furthermore, from (\ref{2Dycks}), it is 
clear that the critical behavior of the ASEP is a combination of that for two adsorbing Dyck walks where the second-order lines correspond to the adsorbtion/desorbtion transition for one of the walks and the first-order line to a co-operative transition not seen for a single walk.  These results are in perfect accordance with those recently derived in \cite{Brak1, Brak2} for the canonical ensemble.  

\section{Comparison with parallel dynamics}
It is interesting to compare  the grand-canonical generating function in (\ref{GCASEPz})
with that describing an ASEP with {\it parallel} dynamics, which was derived in \cite{EP}
using recursion relations rather than a direct summation
\begin{equation}
\fl
 \mathcal{Z}_p (z) =\frac{
 \alpha\beta (1+pz) \left[2(1-p)(\alpha\beta-p^2 z)
    -\alpha\beta b^2(1-p z)
    -\alpha\beta b^2\sqrt{(1+p z)^2-4z}\,\right]}
 {2p^4(1-\beta)(1-\alpha)(z-z_{hd})(z-z_{ld}) }\, ,
 \label{Thfinal}
\end{equation}
where 
 \begin{eqnarray}
b^2&=& \frac{p}{\alpha \beta}\left[ (1-p)-(1-\alpha)(1-\beta)\right]
 \label{bsq}
 \end{eqnarray}
and
\begin{eqnarray}
z_{ld} &=& {\alpha(p-\alpha) \over p^2(1-\alpha)} \\
z_{hd} &=& {\beta(p-\beta) \over  p^2(1-\beta)}\, 
\end{eqnarray}
are the poles which govern the low density and high density phases
respectively.
With a parallel update all possible particle moves
into empty spaces are carried
out simultaneously with probability $p$ at each time step, rather than the random sequential 
updates implicity used elsewhere in the discussion. 
The parallel update ASEP is a special case of the Nagel-Schreckenberg \cite{traffic}
model for traffic flow
and the transitions in the model can be interpreted as jamming transitions for the traffic
in a one dimensional flow. 
The phase diagram has a similar topology,
and similar phases, to the random sequential update ASEP \cite{EP,dGN}, 
but the second order lines
now lie at $\alpha,\, \beta=1 - \sqrt{1-p}$. 

It is clear from the definition of the dynamics that substituting
$\alpha = p \, \tilde \alpha$, $\beta = p \, \tilde \beta$ and taking the limit
$p \rightarrow 0$ will recover random sequential sequential updates.
If we take this limit in  (\ref{Thfinal})
we find
\begin{equation}
 \mathcal{Z}_{p \rightarrow 0} (z) = \tilde\alpha \tilde\beta {\left[1 - \tilde \alpha - \tilde \beta + 2 \tilde \alpha \tilde \beta + (1- \tilde \alpha -\tilde \beta ) \sqrt{1 - 4 z}  \right] 
\over (z - \tilde \alpha(1 - \tilde \alpha) )( z - \tilde \beta ( 1 - \tilde \beta) )} + O ( p)
\end{equation}
where the leading term is just a rationalized form of the grand-canonical generating 
function derived earlier in (\ref{GCASEPz}). The limiting process for obtaining random sequential 
dynamics from parallel dynamics can thus be carried out at the level of the grand-canonical generating
function.

It is also noteworthy that it is possible to write
\begin{equation}
\mathcal{Z}_p (z) = 
{ \alpha \beta ( 1 - x_{-}(z)^2/p ) \over (\alpha  - x_{-} (z) ) ( \beta - x_{-} (z) ) }
\label{PGCASEPz}
\end{equation}
where $x_{-} (z)$ is the negative root of 
\begin{equation}
z = {x (p- x) \over p^2 (1- x)} \, . 
\end{equation}
The $x(z)$ appearing in the grand-canonical generating function for random sequential dynamics (\ref{GCASEP}) is one of the roots of $z = x ( 1 - x)$, so the structure of the parallel grand-canonical generating function above is similar.  In fact, it turns out that the lattice paths described by this latter generating function generalise the Dyck walks of the random-sequential case in a natural way \cite{UsInPrep}.

\section{Discussion}

Taking stock, we have found that the introduction of an initially \textit{ad-hoc} grand-canonical approach to the ASEP leads to an extremely simple, unified 
description of the entirety of the model's steady-state phase behavior, not 
to mention a straightforward confirmation of lattice path equivalences derived 
for the canonical ensemble in \cite{Brak1, Brak2}.  It is, in fact, possible to go further and relate properties of the lattice paths---obtained using standard \emph{equilibrium} statistical mechanical techniques---to observables in the ASEP's nonequilibrium steady state.  For example, density profiles \cite{Brak2} and correlation functions \cite{DEL} have been calculated in this way.  Hence it is evident that the lattice path interpretation of a nonequilibrium partition function can be developed into a useful calculational tool.

In this work we employed the grand-canonical ensemble as a mathematical device.  It is, in fact, possible to give physical meaning to this ensemble if one introduces processes that allow the size of the system to fluctuate at a rate controlled by the fugacity $z$.  A model with such processes was investigated in \cite{Heilmann}.  In that work, the $N$-site chain is thought of as a `cluster' embedded in a larger system of particles that can attach to and detach from the cluster at its boundaries at various rates.  For a particular choice of these rates, in which $z$ is the ratio of the attachment to detachment rate, the normalisation given in \cite{Heilmann} corresponds to our expression (\ref{GCASEPz}).  If particles attach at a rate greater than the current of particles through the cluster, the size of the cluster explodes to infinity, a phenomenon that explains the divergence of the grand-canonical partition function at a critical value of $z$ equal to the internal current.  In \cite{Heilmann}, an analogy was made to a gas-liquid transition.

Given the compactness of the grand-canonical normalisations for the ASEP with both random-sequential (\ref{GCASEPz}) and parallel dynamics (\ref{PGCASEPz}), one might suggest that the grand-canonical approach might be of general utility in establishing equivalences between nonequilibrium and equilibrium steady states.  One way in which this hypothesis might be given further weight would be to derive the analogous quantity for the Partially Asymmetric Exclusion Process (PASEP) \cite{Sas1,BECE,Sas2} in which hops to the left at a rate $q$ are also entertained.  This model contains an additional phase transition at $q=1$ which is of a different nature to the boundary-driven transitions shown in the ASEP phase diagram, Fig.~2.  It may well be that a grand-canonical partition function for the PASEP would shed additional light onto this transition.

Finally we remark that the simple form of (\ref{GCASEPz}), which required 
previously known results as input, is somewhat tantalizing.  In particular 
it hints at the possibility that it might be directly calculable from purely 
physical principles.  This would be an avenue worth pursuing since, as noted 
in the introduction, partition functions appear to have some relevance to 
nonequilibrium steady states, yet there are as yet no systematic approaches 
for obtaining them.

%%%%%%%%%%% ACKNOWLEDG(E)MENTS %%%%%%%%%%%

\section{Acknowledgements}

WJ and DAJ were partially supported by EC network grant {\it
HPRN-CT-1999-00161}. RAB is an EPSRC Postdoctoral Research Fellow supported 
by grant {\it GR/R44768}. We would like to thank the referee for bringing
\cite{Heilmann} to our attention.

\vspace{3.75ex}

%%%%%%%%%%%%%%% REFERENCES %%%%%%%%%%%%%%% 

% bibtex output included and modified to include an endnote and some 
% grouping

\end{document}